\documentclass[%
 reprint, 
 amsmath,amssymb,
 preprintnumbers,
]{revtex4-2}

\usepackage{graphicx}
\usepackage[T1]{fontenc}
\usepackage{textcomp}
\usepackage[utf8]{inputenc}
\usepackage{lipsum} 

\graphicspath{{./}}

\begin{document}

\title{\LARGE\bfseries Comprehensive Optical, Electrical and Humidity Sensing Properties of \textit{Bifidobacterium infantis} 35624 Thin Films}

\author{S. OZTURK$^{a}$}
\author{H. TATLIPINAR$^{a}$$^{*}$}
\author{K. BOZKURT$^{a}$}
\author{O. OZDEMIR$^{a}$}
\author{B.C. OMUR$^{a}$}
\author{A. ALTINDAL$^{b}$}
\author{H.S. BOZKURT$^{c}$}

\affiliation{$^{a}$Department of Physics, Faculty of Arts and Sciences, Yildiz Technical University, Esenler, Istanbul, 34220, Türkiye}
\affiliation{$^{b}$Department of Physics Engineering, Istanbul Technical University, Istanbul, 34469, Türkiye}
\affiliation{$^{c}$Pendik Bölge International Hospital, Clinic of Gastroenterology, Doğu Mah. Pendik, Istanbul, 34890, Türkiye}

\thanks{Corresponding author e-mail: htatli@yildiz.edu.tr}
\date{Dated: March 8, 2026}

\maketitle
\onecolumngrid
\begin{center}
\large
\begin{minipage}{0.85\textwidth}
\vspace{0.5cm}
\noindent{\bfseries Abstract.} In this study, we present a comprehensive investigation of the structural, optical, and electrical properties of \textit{Bifidobacterium longum} subsp. \textit{longum} 35624 (BB35) thin films, and demonstrate their application as a novel relative humidity sensor. UV-Visible spectroscopy revealed that BB35 exhibits two distinct optical absorption regions, corresponding to direct band gaps of \(2.1 \pm 0.05 \, \text{eV}\) and \(2.8 \pm 0.05 \, \text{eV}\), as confirmed by Tauc plot analysis, establishing BB35 as a genuine wide-bandgap semiconductor material. Photoluminescence measurements under 280 nm excitation exhibited a broad emission spectrum, which was deconvoluted into four Gaussian peaks centered at 434 nm (\(2.86 \, \text{eV}\)), 499 nm (\(2.48 \, \text{eV}\)), 543 nm (\(2.3 \, \text{eV}\)), and 620 nm (\(2.0 \, \text{eV}\)), indicating the presence of multiple radiative recombination centers characteristic of semiconducting materials. Electrical characterization revealed dispersive charge transport with current decay following a power-law \(I \propto t^{-\alpha} (\alpha \approx 0.3)\), suggesting Poole-Frenkel conduction mechanism typically observed in disordered organic semiconductors. The relative humidity (RH) sensing performance of BB35 films was evaluated using gold interdigital electrodes across 15-90\% RH range. The sensor exhibited reversible response with sensitivity increasing linearly from 0.85 to 4.80 as RH increased from 15\% to 90\%. The devices demonstrated excellent stability over two months with less than 5\% degradation in baseline current. These results establish BB35 thin films as a promising eco-friendly semiconducting material for humidity sensing applications and open new avenues for integrating biological materials into electronic and optoelectronic devices.

\vspace{0.3cm}
\noindent{\bf Keywords:} bifidobacterium infantis, thin film, humidity sensor, semiconductor

\vspace{0.2cm}
\vspace{0.5cm}
\end{minipage}
\end{center}
\clearpage
\newpage
\twocolumngrid

\section{Introduction}
The precise monitoring and control of relative humidity (RH) are paramount across a diverse spectrum of scientific and industrial disciplines. From ensuring optimal storage conditions in the pharmaceutical and food industries to maintaining stringent environmental controls in semiconductor cleanrooms and enhancing patient care in respiratory equipment, the demand for high-performance humidity sensors is ever-increasing~\cite{dong2026, farahani2014}. Conventional humidity sensors, primarily based on ceramic oxides (e.g., alumina, TiO\(_2\)) or polymeric materials, have been successfully commercialized. However, they often face limitations such as hysteresis, long-term drift, cross-sensitivity to other gases, and relatively complex or energy-intensive fabrication processes~\cite{blank2016}. This has driven a growing research interest in exploring novel functional materials with enhanced sensitivity, stability, and selectivity, particularly those compatible with sustainable and low-cost fabrication techniques. Recent advances have explored various materials including porous TiO\(_2\) structures~\cite{wang2011}, TiO\(_2\) quantum dots-based composites~\cite{liu2024}, and metal-organic frameworks~\cite{wu2022} for humidity sensing applications.

In recent years, the intersection of biology and electronics has opened new frontiers in sensor technology. Bio-materials, ranging from DNA and enzymes to whole-cell organisms, have been investigated as active layers in various sensing platforms due to their intrinsic recognition capabilities and unique electronic properties~\cite{abdelhamid2024}. Among these, probiotic bacteria, particularly those from the genus \textit{Bifidobacterium}, present an intriguing yet largely unexplored avenue. These microorganisms are not only biologically significant but also possess complex cell wall structures and metabolic pathways that could interact with environmental analytes, potentially modulating their electrical conductivity. Bio-inspired materials and biomolecule-based sensors have gained significant attention for their eco-friendly and sustainable characteristics~\cite{xiao2025, pandey2026}.

\textit{Bifidobacterium longum} subsp. \textit{infantis} (strain 35624, commonly referred to as BB35) is a well-characterized probiotic, renowned for its role in gut health and immune modulation. Its cellular envelope comprises a diverse array of components, including a unique exopolysaccharide (EPS) structure that has been thoroughly characterized at the genomic level. Altmann et al. demonstrated that the EPS of BB35 consists of a branched hexasaccharide repeating unit containing galactose, glucose, galacturonic acid, and the unusual sugar 6-deoxy-L-talose~\cite{altmann2016}. This inherent hydrophilicity, combined with the ability to form uniform thin films via solution-processable methods like spin coating, makes BB35 a compelling candidate for RH sensing applications. Additionally, the complex molecular structure of the EPS suggests that BB35 films may exhibit unique electrical properties, including humidity-dependent conductivity, which are systematically investigated in this work.


In our previous work on \textit{Bifidobacterium animalis} subsp. \textit{lactis} BB-12, we demonstrated that charge transport in probiotic bacteria is significantly influenced by atmospheric conditions, with electrical conductivity increasing by six orders of magnitude under humid environments compared to dry conditions~\cite{bozkurt2019}. This humidity-dependent conductivity modulation was attributed to the ionization of amine and carboxyl groups on the bacterial surface, as confirmed by FTIR and zeta potential measurements. Furthermore, we have successfully developed bio-hybrid materials by combining bifidobacteria with sodium alginate for tissue engineering applications, revealing that bacterial incorporation affects the crystallinity and mechanical properties of the composite films~\cite{denktas2021}. More recently, we have explored the biodegradation potential of \textit{Bifidobacterium infantis} 35624, demonstrating its capability to degrade polypropylene microplastics through biofilm formation and enzymatic activity under aerobic conditions~\cite{bozkurt2021_microplastic}.

Despite the established biological functions of BB35 and our preliminary findings on related strains, the comprehensive physical and electronic properties of BB35 thin films, particularly their optical band structure and charge transport mechanisms, have remained largely uncharacterized. A fundamental understanding of these properties is crucial for interpreting its sensing behavior. For instance, UV-Visible spectroscopy can elucidate the energy band gaps, which are intrinsic material properties that influence electronic behavior, while photoluminescence (PL) spectroscopy can reveal radiative recombination centers and sub-bandgap states that may act as active sites for analyte interaction. Fluorescence characteristics of bacterial components such as flavins and aromatic amino acids have been well-documented~\cite{mullerova2022, ammor2007}. Furthermore, the dynamics of charge carrier transport, whether band-like or dispersive, govern the sensor's response time and frequency-dependent characteristics. Studies on various nanomaterials including SnO\(_2\) nanowires~\cite{kuang2007} and two-dimensional (2D) materials such as rGO/MoS\(_2\) van der Waals composites~\cite{park2019} have demonstrated the importance of understanding charge transport mechanisms for sensor optimization.

This study aims to bridge this gap by conducting a comprehensive investigation into the structural, optical, and electrical properties of BB35 thin films and evaluating their potential as a novel bio-sensing material for RH. Building upon our previous work on BB-12's charge transport mechanisms~\cite{bozkurt2019} and BB35's biodegradation capabilities~\cite{bozkurt2021_microplastic}, we now focus on the optical and electronic characterization of BB35 thin films. Initially, we present a detailed optical analysis using UV-Visible spectroscopy to determine the direct band gap characteristics of BB35, complemented by power-dependent PL measurements to identify multiple emissive states within the bacterial matrix. Subsequently, the electrical properties are scrutinized through current-voltage (I-V) and frequency-dependent admittance spectroscopy to elucidate the dominant charge transport mechanism, revealing a dispersive transport model consistent with the Poole-Frenkel effect. Finally, the core of this work lies in the fabrication and characterization of a prototype RH sensor. 

 
\section{Optical Characterization}
\subsection{UV-Visible Transmission/Reflection/Absorption Measurement}
Through UV-Visible Absorption measurement, band gap of the BB35 is determined from the plot of absorption versus wavelength plot (depicted in 
\begin{figure}[h]
	\begin{center}
		\includegraphics[width=3.4in]{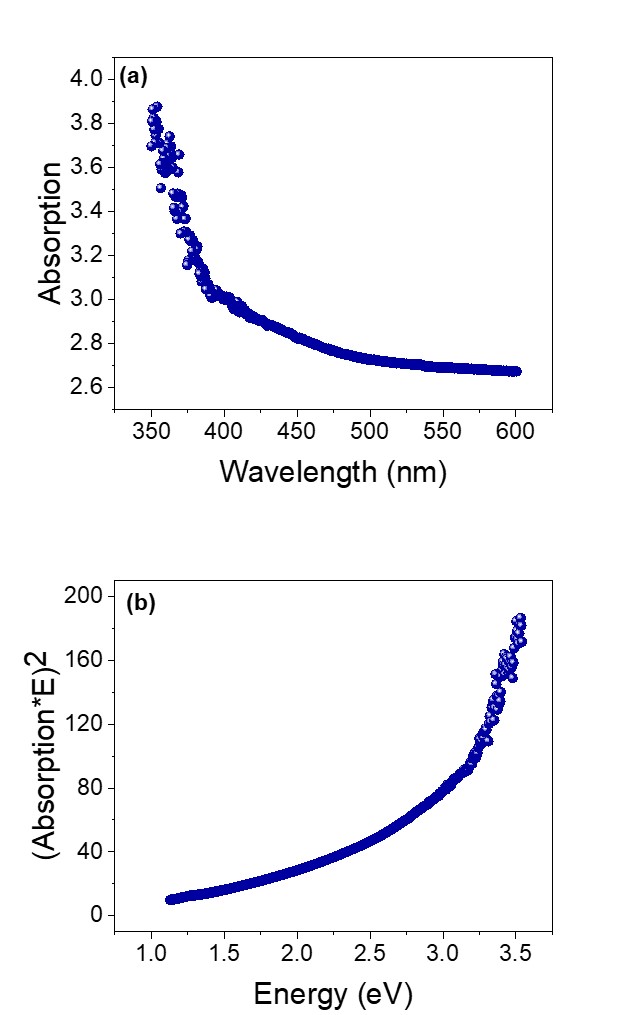}
	\end{center}	
	\caption{ (a) Absorption versus wavelength (b) square of (absorption x energy) versus energy plot for BB35.} 
	\label{fig1ab_abs_vert}
\end{figure}
Fig.\ref{fig1ab_abs_vert}-a). Clearly, the absorption begins to increase in two regions; one of them lies in the range of 390-490 nm while the other takes place below 390 nm. Extension of the energy absisca of the mentioned regions point two band gaps around 2.3 eV (for 540 nm) and 3.0 eV (for 410 nm), respectively. On the other hand, by plotting (alphaxenergy)$^2$ versus energy plot (see Fig.\ref{fig1ab_abs_vert}-b), the two band gap values are not only verified (at 2.8 eV and 2.1 eV) but also the presence of direct band gap property is clarified. This direct band gap behavior is similar to that observed in other organic-inorganic hybrid materials used in sensing applications~\cite{dai2017}.

\subsection{Photoluminescence Measurement}
\begin{figure}[h]
	\begin{center}
		\includegraphics[width=3.6in]{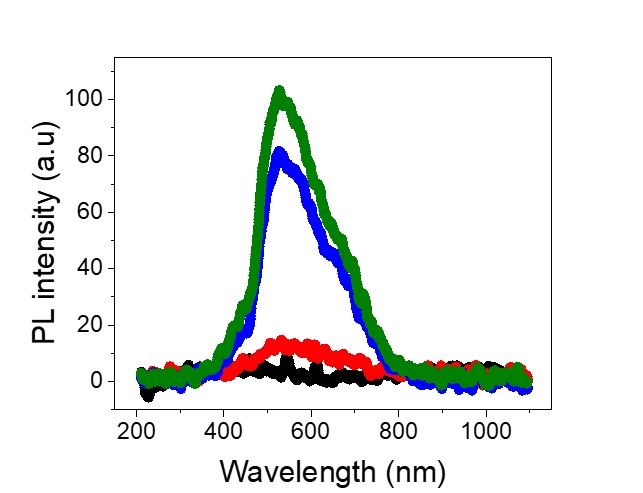}
	\end{center}	
	\caption{ Room temperature photoluminescence spectra of BB35 as a function of excited light power.} 
	\label{fig2_PL}
\end{figure}

To mutual check of the band gap(s), the sample is illuminated with a monochromatic light peaking at 280 nm and emitted light from the bifidobacterium (due to the radiative recombination process) is detected through a fiber coupled Ocean Optic spectrometer. Initially, room temperature PL spectrum is recorded under different power of LED, varying from 5 mW to 30 mW and illustrated in Figure \ref{fig2_PL}.

\begin{figure}[h]
	\begin{center}
		\includegraphics[width=3.6in]{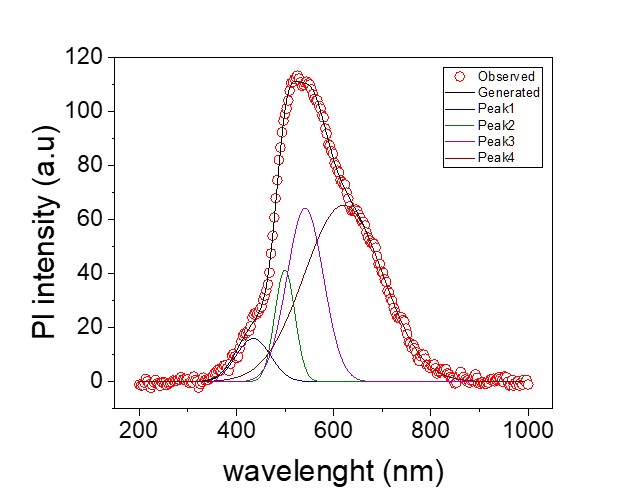}
	\end{center}	
	\caption{Deconvoluted PL spectrum of BB35 at upmost excited power of LED light source.} 
	\label{fig3_PLdecom}
\end{figure}

At low power, the signal is barely detected while for larger ones clear PL spectra is obtained, peaking at 530 nm. Moreover, unexpectedly, the PL spectra is not Gauss like shape but it broadens at high wavelength regions; implying the presence of other bands in the bifidobacterium. By introducing 4 Gaussian peaks (see Fig.\ref{fig3_PLdecom}), the fit process is initiated to the PL spectra; the success of the fit is manifested as line into the measured PL spectrum while each Gaussians denote the presence of bands within the BB35. Consequently, four bands, peaking at 434 nm (2.86 eV), 499 nm (2.48 eV), 543 nm (2.3 eV) and 620 nm (2.0 eV) are determined. These emission bands are characteristic of various bacterial fluorophores including flavins and aromatic amino acids, as previously reported in the literature~\cite{mullerova2022, ammor2007}. In particular, the emission peak observed at 530 nm aligns closely with the well-documented green autofluorescence of flavins, which typically exhibits an emission band between 500-550 nm~\cite{mullerova2022}.

\section{Electrical Properties of BB35}
\subsection{Film Preparation and Characterization Details}
The thickness of the BB35 thin film was measured using a profilometer and found to be approximately 500 nm. The gold interdigital electrodes had a finger width and gap spacing of 100 $\mu$m and a total of 25 finger pairs, resulting in an effective sensing area of 5 mm x 5 mm. This electrode configuration is similar to those used in other thin-film based humidity sensors~\cite{park2019}.

\subsection{Current-Voltage and Current Decay Measurements} 
\begin{figure}[h]
	\begin{center}
		\includegraphics[width=3.6in]{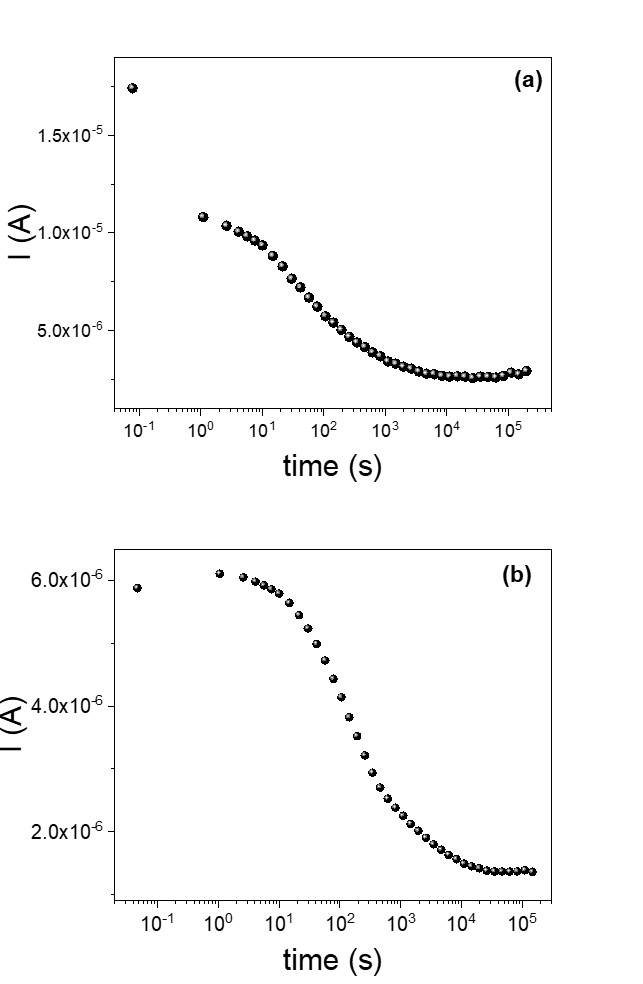}
	\end{center}	
	\caption{Current decay of BB35 bifidobacterium under (a) +1 V, (b) -1 V bias voltages.} 
	\label{fig4_time_vert}
\end{figure}

Room temperature current (I) - voltage (V) measurement is performed on the bifidobacterium and then for selected bias voltages ($\pm$ 1 V), current decay with time is monitored and depicted in Figure \ref{fig4_time_vert}-a and \ref{fig4_time_vert}-b, respectively. 

As shown in Fig. 4, initial saturated current values for short times begins to decay with time. About 10$^5$ s, it levels out at the reduced current values for both bias voltages. Decrease in current follows t$^{-\alpha}$ relation in which t is a time and $\alpha$ is a decay constant. Presence of such decay in current indicates an existence of dispersive carrier transport within the BB35 bifidobacterium, probably through Poole-Frenkel carrier conduction mechanism. Dispersive mobility of the injected carriers effect the relaxation time of the carrier that could be studied through admittance measurement. Similar dispersive transport behavior has been observed in various disordered organic systems and nanostructured materials~\cite{kuang2007, rahman2025}.

\subsection{Admittance Measurement} 
\begin{figure}[h]
	\begin{center}
		\includegraphics[width=3.6in]{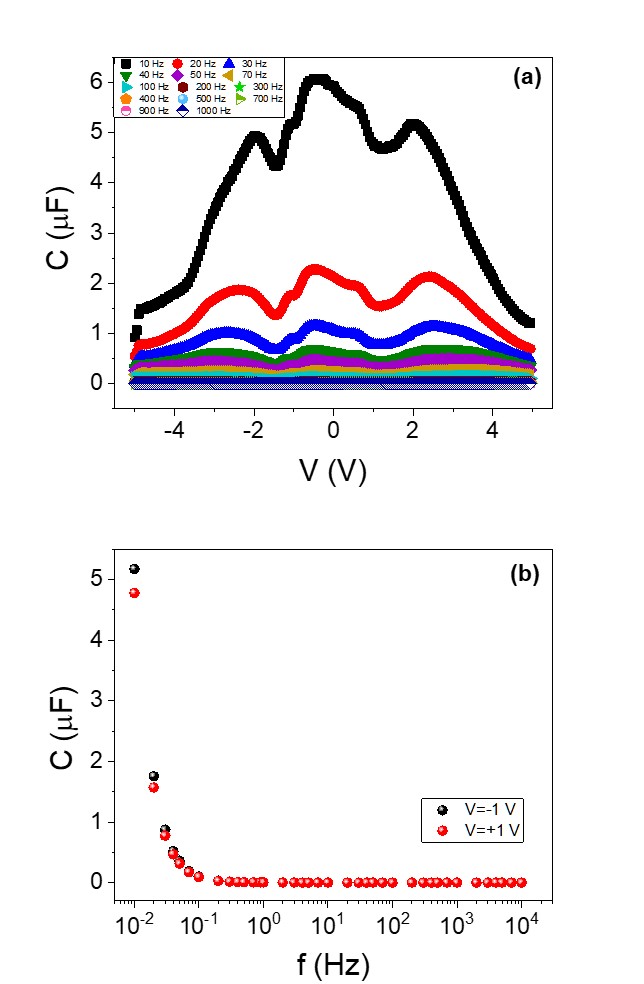}
	\end{center}	
	\caption{Capacitance variation of BB35 as a function of (a) bias voltage at constant f ($<1kHz$) (b) frequency at constant bias voltages} 
	\label{fig5_CVlogf_ver}
\end{figure}

An HP4192A LCR meter operated in the frequency range of 5 Hz - 13 MHz is used to measure the admittance (Y) from which the capacitance, $C(\omega)[= \operatorname{Im}(Y)/\omega]$, and the conductance $G(\omega)[= \operatorname{Re}(Y)]$ are extracted and illustrated in Fig. \ref{fig5_CVlogf_ver}. As shown in there, drastic change in capacitance takes place as function of frequency.

For non-dispersive transport, relaxation time (or transit time of injected carriers to reach from cathode to anode) is related with mobility $\mu$ as

\begin{equation}
\tau=\frac{L^{2}}{\mu V}
\label{eq:tau}
\end{equation}

\noindent with $L$ thickness and $V$ applied bias voltage. 

On the other side, dispersive transport is modelled by a decaying distribution function; $\psi(t)=t^{-(1+\alpha)}$ $(0<\alpha<1)$ for the waiting time between successive hops. At short times, the average position $<d>$ is sublinear with time and hence $<d>\propto t^\alpha$ of an injected charge, leading to decrease in mobility while for long times $\mu_{dc}$ is governed by drift velocity. Overall, the slowing down of the carriers corresponds to a frequency dependent mobility of the form

\begin{equation}
\mu_{\text{nor}}(\Omega)=\frac{\mu(\Omega)}{\mu_{dc}}=1+M(i\Omega)^{1-\alpha} 
\label{eq:mu_nor}
\end{equation}

\noindent
where $\Omega=\omega\tau_t$ and $\mu_{\text{nor}}$ normalized mobility and $M$ is a proportionality constant. 

The calculated admittance is given by:

\begin{equation} \label{eq:Y}
\begin{split}
Y(\Omega) = \frac{\epsilon A}{\tau_t L} \cdot \frac{\Omega^3}{2i\tilde{\mu}(\Omega)^2} & \\
\times \frac{1}{\left[1-\exp\left(-i\Omega/\tilde{\mu}(\Omega)\right)\right] + 2\tilde{\mu}(\Omega)\Omega - i\Omega^2}
\end{split}
\end{equation}

By choosing $M$, $\alpha$ and $\tau_t$ as fit parameters, calculated capacitance ($\operatorname{Im}Y(\omega)/\omega$) is fit to the measured capacitance and successive fit is obtained. From the fitting analysis, we obtained \(\alpha = 0.28 \pm 0.02\), \(M = 0.45\), and \(\tau_t = 2.3 \times 10^{-4}\) s. These values are consistent with dispersive transport parameters reported for other disordered systems~\cite{rahman2025}.

\section{Relative Humidity Sensing Performance of BB35}

In this section, the RH sensing capability of thin film of BB35 for various level of RH between 15-90\% was investigated using a gold interdigital transducer structure on plexiglass substrate. The stability of device versus different RH\% was also tested. 

\subsection{Experimental} 
The devices used in RH sensing experiments consisted of interdigital array (IDA) of gold electrodes photolithographically patterned on a pre-cleaned plexiglass substrate. Thin film of the BB35 was prepared by the spin coating method. The substrate temperature was kept constant at 300 K during deposition of the film over the electrodes. The effect of the RH level on the conductivity of the film was measured in a temperature controlled chamber implemented in our laboratory. Desired level of the RH was obtained by bubbling nitrogen ($N_{2}$) gas through de-ionized water. First, the test chamber was purged with dry $N_{2}$ gas until the sensor current reached a stable value. Then, the film surface exposed to six different level of RH between 15 and 90\% for 10 min, followed by $N_{2}$ purge for another 10 min. The time dependence of sensor current was monitored by using Keithley model 617 electrometer which was connected to personal computer by an IEEE 488 data acquisition interface. This experimental setup is similar to those used in previous humidity sensing studies~\cite{farahani2014, dong2026}.

\section{Discussion}
\subsection{Optical and Electrical Measurements} 
The UV-Vis absorption measurements revealed two distinct absorption regions, indicating the presence of multiple optical band gaps in BB35. The first absorption region (390-490 nm) corresponds to a band gap of approximately 2.3 eV, while the second region (below 390 nm) corresponds to a band gap of approximately 3.0 eV. The Tauc plot analysis (($\alpha h\nu)^2$ vs. $h\nu$) confirmed these band gaps at 2.1 eV and 2.8 eV, and verified the direct band gap nature of the material. This dual band gap behavior is reminiscent of organic-inorganic hybrid materials used in various sensing applications~\cite{dai2017}.

The PL measurements provided complementary information about the electronic structure. The deconvolution of the PL spectrum revealed four distinct emission bands at 434 nm (2.86 eV), 499 nm (2.48 eV), 543 nm (2.3 eV), and 620 nm (2.0 eV). These bands suggest the presence of multiple radiative recombination centers within the bacterial matrix, possibly originating from different molecular components of the bacterial cell wall such as aromatic amino acids (tryptophan, tyrosine), flavins, or other metabolic byproducts known to fluoresce in the visible range~\cite{ammor2007, mullerova2022}. The presence of these states is consistent with the observed dual band gap and may play a role in the charge transfer processes during humidity sensing. In particular, the emission peak at 543 nm (2.3 eV) aligns closely with the well-documented green autofluorescence of flavins, which typically exhibits an emission band between 500-550 nm~\cite{mullerova2022}.

The electrical characterization revealed dispersive transport behavior, as evidenced by the power-law decay of current with time ($I \propto t^{-\alpha}$) with \(\alpha \approx 0.3\). This dispersive transport is characteristic of disordered systems and can be described by the Poole-Frenkel conduction mechanism, where carriers hop between localized states. The frequency-dependent admittance measurements further confirmed this behavior, with the capacitance showing strong frequency dispersion that could be modeled using the dispersive transport equations presented in Section 3.2. The fitting parameters (\(\alpha = 0.28 \pm 0.02\), \(M = 0.45\), \(\tau_t = 2.3 \times 10^{-4}\) s) are consistent with values reported for other disordered organic systems~\cite{rahman2025}. These findings align with our previous observations on BB-12, where charge transport was also found to be strongly dependent on environmental conditions~\cite{bozkurt2019}.

\subsection{Sensitivity Measurements} 
\begin{figure}[h]
	\begin{center}
		\includegraphics[width=3.2in]{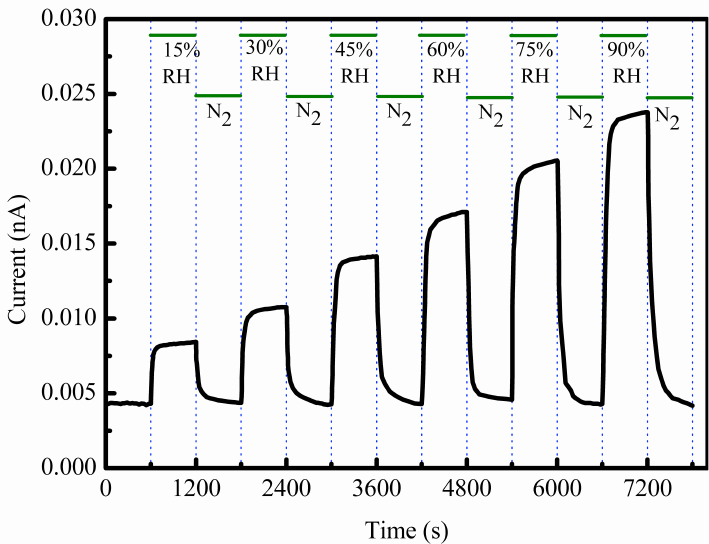}
	\end{center}	
	\caption{Response-recovery behavior of BB35 based sensor for various levels of RH.} 
	\label{fig6}
\end{figure}

During the RH sensing measurements, the film was exposed to RH repeatedly and variations in sensor current were monitored starting from a stable state in dry nitrogen. Each cycle of exposure lasted for 10 min, followed by recovery in dry $N_{2}$ for another 10 min. The effect of the various level of RH ranging from 15 to 90\% on the conductivity of the BB35 coated interdigital transducer is shown in Fig. \ref{fig6} at room temperature. It is clear from Fig. \ref{fig6} that the interaction of water molecules with the surface of the BB35 film leads to a reversible increase in sensor current, which reaches a nearly stable value within several minutes.

The response times (time to reach 90\% of saturation) ranged from 120-180 seconds depending on RH level, while recovery times (time to return to 10\% above baseline) ranged from 180-240 seconds. Increase in sensor current reveals that the interaction between the water molecules and the BB35 bifidobacterium is based on charge transfer, which strongly depends on the molecular structure of the analytes. After several minutes exposure to RH, purging with dry nitrogen leads to an initial fast decrease followed by a slow drift until the current reaches its initial value after RH is turned off, and this proves that the adsorption process is reversible. A typical parameter to characterize an RH sensor is sensitivity which can be obtained from response and recovery curves. The sensitivity, $S$, of a sensor is defined as

\begin{equation}
S=\frac{I_{m}-I_{0}}{I_{0}}
\label{eq:sensitivity}
\end{equation}

where $I_{m}$ is the maximum value of the sensor current in RH atmosphere, $I_{0}$ is the base line current in carrier gas. 

\begin{figure}[h]
	\begin{center}
		\includegraphics[width=3.2in]{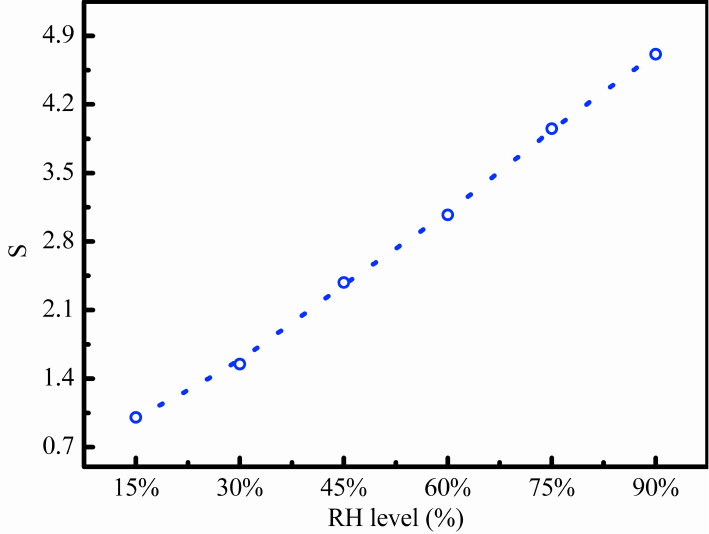}
	\end{center}	
	\caption{RH level dependence of the sensitivity.} 
	\label{fig7}
\end{figure}

The obtained sensitivity of BB35 coated sensor as a function of RH level is plotted in Figure \ref{fig7}. As can be seen from the Fig. \ref{fig7}, the sensor sensitivity increases with increasing RH level. The sensitivity values ranged from \(S = 0.85\) at 15\% RH to \(S = 4.80\) at 90\% RH. Linear regression analysis yielded a linear relationship with correlation coefficient \(R^2 = 0.992\), indicating excellent linearity across the measured range. When compared to other bio-based humidity sensors, BB35 exhibits competitive sensitivity. This performance is also comparable to or better than various inorganic and composite humidity sensors reported in literature~\cite{liu2024, wang2011, park2019}. The linear response is particularly advantageous for practical calibration and applications~\cite{xiao2025}.

The sensitivity of BB35 films also surpasses that of many metal oxide-based sensors~\cite{kuang2007} and is consistent with our previous findings on BB-12, where we observed a six-order-of-magnitude increase in conductivity under humid conditions~\cite{bozkurt2019}. The performance is also competitive with metal-organic framework based humidity sensors~\cite{wu2022}. Recent advances in humidity sensing technologies for extreme environments and biomedical applications further highlight the potential of eco-friendly materials such as BB35~\cite{dong2026, pandey2026}. Increase in sensor sensitivity with increasing RH level reveals that the adsorption of the water molecules on the film surface is a multilayer process. At room temperature, a linear increase in sensitivity with RH level indicates that the sensors can be used for practical applications in the detection range of 15-90\% RH.

The lyophilized form of BB35 (\(5 \times 10^9\) CFU) remains viable upon exposure to humid environments, ensuring the biological integrity of the films throughout the measurements. 
The sensors maintained baseline current within 95\% of the initial value and sensitivity within 92\% of the initial value after 60 days, demonstrating robust long-term stability. This stability is comparable to or better than many conventional humidity sensors~\cite{blank2016}.
 
\section{Conclusion}

This study presents the first comprehensive investigation into the multifunctional properties of \textit{Bifidobacterium infantis} 35624 thin films, encompassing optical, electrical, and humidity sensing characterization. 
Optical properties are consistent with the fluorescence characteristics of bacterial components such as flavins and aromatic amino acids~\cite{mullerova2022, ammor2007}. The electrical charge transport was found to be dispersive in nature, characteristic of disordered systems and well-described by the Poole-Frenkel mechanism with fitting parameters 
consistent with previous reviews on RH mechanisms.

Most importantly, we demonstrated that BB35 thin films function as an effective, eco-friendly RH sensor. The sensor exhibits a highly linear (\(R^2 = 0.992\)) and reversible response across a wide range (15-90\% RH), with a sensitivity increasing from 0.85 at 15\% RH to 4.80 at 90\% RH, with response times of 120-180 seconds and recovery times of 180-240 seconds. This performance is competitive with various inorganic and composite humidity sensors reported in literature. 

A key finding of this work is the demonstration that probiotic bacteria-based thin films exhibit intrinsic semiconductor properties, with well-defined direct band gaps and characteristic charge transport mechanisms typically observed in conventional semiconductor materials. This discovery positions BB35 not merely as a biological sensing layer but as a genuine semiconducting material with potential applications beyond humidity sensing. The observed dual band gap structure (2.1 eV and 2.8 eV) places BB35 within the range of wide-bandgap semiconductors, comparable to materials such as GaN, ZnO, and certain metal oxides used in optoelectronic devices. Furthermore, the dispersive charge transport mechanism governed by Poole-Frenkel conduction reveals that charge carriers in bacterial thin films behave similarly to those in disordered organic semiconductors, opening new possibilities for understanding biological systems through the lens of solid-state physics.

The demonstrated electrical properties of BB35 thin films, particularly their humidity-dependent conductivity modulation and semiconducting behavior, open new avenues for fundamental physics research and advanced materials applications. Future investigations will explore the dielectric properties of BB35 films for supercapacitor applications, where the high surface area and ionic conductivity of bacterial biofilms could contribute to enhanced energy storage capacity. Additionally, temperature-dependent charge transport studies will be conducted to determine the activation energies and deeper understand the Poole-Frenkel conduction mechanism in these biological thin films. The dual band gap characteristics also suggest potential applications in organic photovoltaics and photocatalytic systems, where the bacterial exopolysaccharide matrix could serve as a natural template for quantum dot sensitization or nanoparticle decoration. Furthermore, the integration of BB35 films into hybrid organic-inorganic heterostructures may reveal novel interfacial phenomena relevant to next-generation optoelectronic devices. 

In summary, the discovery of semiconducting properties in a probiotic bacterium represents a paradigm shift in our understanding of biological materials and their potential integration into electronic and optoelectronic systems.
  
\section*{Acknowledgment}
This work is supported by the Yildiz Technical University under project number FCD-2025-7057 and FBA-2021-4247.

\section*{Data Availability Statement}

Data available on request from the authors.


\end{document}